\documentclass[prd,aps,showpacs,nofootinbib,tightenlines]{revtex4}
\usepackage {amsmath}
\usepackage {amssymb}
\usepackage {epsfig}
\usepackage {graphicx}
\begin{document}

\newcommand{\beq}{\begin{eqnarray}}
\newcommand{\eeq}{\end{eqnarray}}
\newcommand{\non}{\nonumber\\ }

\title{Quasi-two-body decays $B_{(s)}\to P\rho\to P\pi\pi$ in the perturbative QCD approach}

\author{Ya Li$^1$}               \email{liyakelly@163.com}
\author{Ai-Jun Ma$^1$}         \email{theoma@163.com}
\author{Wen-Fei Wang$^2$ \footnote{Corresponding Author}}   \email{wfwang@sxu.edu.cn}
\author{Zhen-Jun Xiao$^{1,3}$}    \email{xiaozhenjun@njnu.edu.cn}

\affiliation{$^1$ Department of Physics and Institute of Theoretical Physics,
                          Nanjing Normal University, Nanjing, Jiangsu 210023, P.R. China}
\affiliation{$^2$ Institute of Theoretical Physics, Shanxi University, Taiyuan, Shanxi 030006, China}
\affiliation{$^3$ Jiangsu Key Laboratory for Numerical Simulation of Large Scale Complex
Systems, Nanjing Normal University, Nanjing, Jiangsu 210023, P.R. China}

\date{\today}

\begin{abstract}
In this work, we calculate the $CP$-averaged branching ratios and the direct $CP$-violating asymmetries of the
quasi-two-body decays  $B_{(s)} \to P (\rho \to) \pi\pi$ by employing the perturbative QCD (PQCD) approach
(here $P$ stands for a light pseudoscalar meson $\pi, K, \eta$  or $\eta^{\prime}$).
The vector current timelike form factor $F_{\pi}$, which contains the final-state interactions between the
pion  pair  in the resonant region associated with the $P$-wave states $\rho(770)$ along with the two-pion
distribution amplitudes,  is employed to describe the interactions between the $\rho$ and the pion pair under
the hypothesis of the conserved vector current.
We found that (a) the PQCD  predictions for the branching ratios and the direct
$CP$-violating asymmetries for most considered $B_{(s)} \to P (\rho \to) \pi\pi$ decays
agree with currently available data within errors,
(b) for ${\cal B}(B \to \pi^0 \rho^0 \to \pi^0(\pi^+ \pi^-)$, the PQCD prediction is
much smaller than the measured one, and
(c) for the $B^+ \to \pi^+(\rho^0\to)\pi^+ \pi^-$ decay mode, there is a negative $CP$ asymmetry
$(-27.5^{+3.0}_{-3.7})\%$, which agrees with other theoretical predictions
but is different in sign from those reported by {\it BABAR} and LHCb Collaborations.
\end{abstract}

\pacs{13.20.He, 13.25.Hw, 13.30.Eg}

\maketitle


\section{Introduction}
Experimental data from different collaborations, like
{\it BABAR}~\cite{BABAR:01,BABAR:02,BABAR:03,BABAR:04,BABAR:05},
Belle~\cite{Belle:01,Belle:02,Belle:03,Belle:04} and LHCb~\cite{LHCb:01,LHCb:02,LHCb:03},
provide valuable information for the three-body hadronic
$B$  meson decays. For these decay modes, both the resonant
and nonresonant contributions may appear, as well as the
possible significant final-state interactions
(FSIs)~\cite{prd89-094013,1512-09284,89-053015}.
Different frameworks have been developed for the study of the three-body hadronic $B$
meson decays, based on the symmetry principles~
\cite{plb564-90,prd72-094031,plb727-136,prd72-075013,prd84-056002,plb726-337,
plb728-579,IJMPA29-1450011,prd91-014029}
or factorization theorems~\cite{FA-01,FA-02,FA-03,TFA-01,TFA-02,prd60-094014,prl83,
npb591,npb606,npb675}.
The QCD-improved factorization (QCDF)~\cite{prl83,npb591,npb606,npb675} has been widely used
in the study of the three-body charmless hadronic $B$ meson
decays~\cite{plb622-207,prd74-114009,B.E:2009th,ST:15,CY01,CY02,CY16}.
In Refs.~\cite{CY02,CY16}, the authors studied the nonresonant
contributions using heavy meson chiral perturbation theory
(HMChPT)~\cite{prd46-1148,prd45-2188,plb280-287} with some modifications and analyzed
the resonant contributions with the isobar model in terms of the usual Breit-Wigner
formalism~\cite{BW-model}.
The perturbative QCD (PQCD) approach based on the $k_T$ factorization
theorem~\cite{Chen:2002th,Chen:2004th} has also
been adopted in Refs.~\cite{Wang-2014a,Wang-2015a,Wang-2016,ly15,ma16}.

As discussed in Refs.~\cite{Chen:2002th,Chen:2004th,Wang-2014a,Wang-2015a}, the hard
$b$-quark decay kernels
containing two virtual gluons at leading order is not important due to the power-suppression.
The contributions from the region, where there is at least one pair of light mesons
having an invariant mass below
$O(\bar\Lambda m_B)$~\cite{Chen:2002th,Chen:2004th}, $\bar\Lambda=m_B-m_b$ being the
$B$ meson and $b$ quark mass difference,
is dominant. It's reasonable that the dynamics associated with the pair of mesons can
be factorized into a two-meson distribution
amplitude $\Phi_{h_1h_2}$~\cite{MP}. As a result, one can describe the typical PQCD
factorization formula for a
$B\to h_1h_2h_3$ decay amplitude as the form of ~\cite{Chen:2002th,Chen:2004th}
 \begin{eqnarray}
\mathcal{A}=\Phi_B\otimes H\otimes \Phi_{h_1h_2}\otimes\Phi_{h_3}.
\end{eqnarray}
With the hard kernel $H$ describes the dynamics of the strong and electroweak
interactions in three-body
hadronic decays in a similar way as the one for the two-body $B\to h_1 h_2$ decays,
the $\Phi_B$ and $\Phi_{h_3}$ are the wave functions for the B meson
and the final-state $h_3$, which absorb the non-perturbative dynamics in the process.
The $\Phi_{h_1h_2}$ is the two-hadron ($h_1$ and $h_2$) distribution amplitude proposed in
Refs.~\cite{MP,MT01,MT02,MT03,MN,Grozin01,Grozin02}, which describes the structure
of the final-state $h_1$-$h_2$ pair.

With the help of the two-pion distribution amplitudes,
quasi-two-body decays $B\to K\rho\to K\pi\pi$,  the subprocesses of the three-body
decays $B\to K\pi\pi$,
have been studied in the Ref.~\cite{Wang-2016} in the PQCD approach utilizing framework discussed
in~\cite{Chen:2002th,Chen:2004th,Wang-2014a,Wang-2015a}. The consistency between
the PQCD predictions and the data
supports the usability of the quasi-two-body framework in Ref.~\cite{Wang-2016}  for the study of
the three-body hadronic $B$ decays.
In this work, we extend the previous studies in Ref.~\cite{Wang-2016} to the
quasi-two-body decays $B\to P\rho\to P\pi\pi$, with the $P$ standing for the light
pseudoscalar mesons, $P=(\pi, K, \eta$ or $\eta^\prime)$, as shown in Fig.~1.
In literature, many works have been done for the decays of $B \to P\rho $
in two-body framework~\cite{npb675,prd60-094014,epjc23275,prd73-074002,ZR01,ZZ:1}
and some of the experimental data could be found in~\cite{BB:01,BB:02,BB:03,BB:04}.
From~\cite{Wang-2016}, we know that
the width of the resonant state $\rho$ and the interactions between the final states
pion pair will show their
effects on the branching ratios especially on the direct $CP$ violations of the
quasi-two-body decays.
We should not neglect these effects in  $B \to P\rho$ decays.
In order to describe the strong interactions between the $P$-wave
resonant state $\rho$ and the final-state pion pair, vector current timelike form
factor $F_\pi$ containing
final-state interactions between pion pair has been employed in Ref.~\cite{Wang-2016}.
Guaranteed by the
Watson theorem~\cite{PR88-1163}, the results from the $\pi$-$\pi$ scattering and $\tau$ decays
for the timelike form factor $F_\pi$ could be borrowed for the study of quasi-two-body $B$
meson decays.
The detailed discussion of $F_\pi$ could be found in~\cite{Wang-2016} and its references.

\begin{figure}[tbp]
\centerline{\epsfxsize=12cm \epsffile{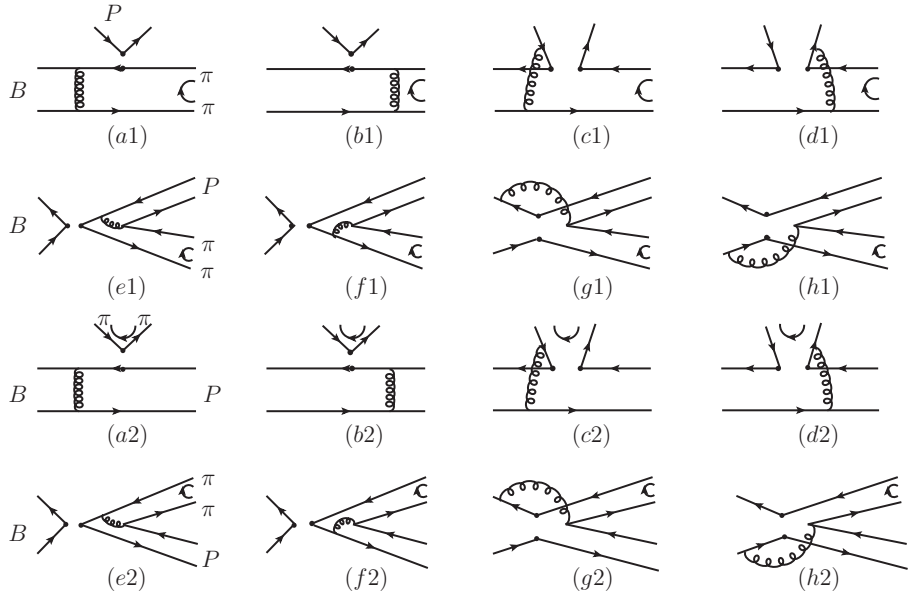}}
\vspace{0.3cm}
\caption{Typical Feynman diagrams for the quasi-two-body decays $B \to P(\rho \to) \pi \pi$,
where $B$ stands for the $B^\pm, B^0$ or $B_s$ meson and $P$ denotes $\pi, K, \eta$ or $\eta^\prime$.
With $\alpha=a$-$d$ and $\beta=e$-$h$, the diagrams ($\alpha$1) for the
$B\to\rho\to\pi\pi$ transition and ($\alpha$2) for the $B\to P$ transition,
as well as the diagrams($\beta$1) and ($\beta$2) for annihilation contributions.}
\label{fig-fig1}
\end{figure}

This paper is organized as follows.
In Sec.~II, we give a brief introduction for the theoretical framework.
The numerical values, some discussions and the conclusions will be given in last two sections.

\section{FRAMEWORK}\label{sec:2}  

For the quasi-two-body $B \to P (\rho \to) \pi\pi$ decays, the weak effective
Hamiltonian can be specified as~\cite{buras96}:
\begin{eqnarray}
 {\cal H}_{eff} &=& \frac{G_{F}}{\sqrt{2}}     \Bigg\{ V_{ub}^{\ast} V_{uq}\Big[     C_{1}({\mu}) O^{u}_{1}({\mu})
                        +  C_{2}({\mu}) O^{u}_{2}({\mu})\Big]   -V_{tb}^{\ast} V_{tq} \Big[{\sum\limits_{i=3}^{10}} C_{i}({\mu}) O_{i}({\mu})
                       \Big ] \Bigg\} + \mbox{H.c.} \;,  \label{eq:hamiltonian01}
\end{eqnarray}
with $q=d, s$, the $C_i(\mu)$($i=1,\ldots,10$) are the Wilson coefficients
and $O_i$  are the local four-quark operators.

We let the pion pair and the final-state $P$ move along the direction of
$n=(1,0,0_{\rm T})$ and $v=(0,1,0_{\rm T})$ in
the light-cone coordinates, respectively. The $B$ meson momentum $p_{B}$,
the total momentum of the pion pair,
$p=p_1+p_2$, and the final-state $P$ momentum $p_3$ are chosen as
\begin{eqnarray}\label{mom-pBpp3}
p_{B}=\frac{m_{B}}{\sqrt2}(1,1,0_{\rm T}),~\quad p=\frac{m_{B}}{\sqrt2}(1,\eta,0_{\rm T}),~\quad
p_3=\frac{m_{B}}{\sqrt2}(0,1-\eta,0_{\rm T}),
\end{eqnarray}
where $m_{B}$ is the mass of $B$ meson, the variable $\eta$ is defined as $\eta=\omega^2/m^2_{B}$,
the invariant mass squared $\omega^2=p^2$.
We define $\zeta=p^+_1/p^+$ as one of the pion pair's momentum fraction,
in terms of which the other kinematic variables of the two pions are expressed as
\begin{eqnarray}
p^-_1=(1-\zeta)\eta\frac{m_{B}}{\sqrt2}, \quad p^+_2=(1-\zeta)\frac{m_{B}}{\sqrt2},
\quad p^-_2=\zeta\eta\frac{m_{B}}{\sqrt2}.
\end{eqnarray}

We employ $x_B, z, x_3$ to denote the momentum fraction of the positive quark in each meson,
$k_{BT}, k_{\rm T}, k_{3{\rm T}}$ stands for the transverse momentum of the positive quark,
respectively.
The momentum $k_B$ of the spectator quark in the $B$ meson, the momentum $k$ for the resonant
state $\rho$
and $k_3$ for the final-state $P$ are of the form of
\begin{eqnarray}
k_{B}&=&\left(0,x_B \frac{m_{B}}{\sqrt2} ,k_{BT}\right),\quad k= \left( \frac{m_{B}}{\sqrt2}z,0,k_{\rm T}\right),\quad
k_3=\left(0,(1-\eta)x_3 \frac{m_B}{\sqrt{2}},k_{3{\rm T}}\right), \label{mom-B-k}
\end{eqnarray}
The momentum fractions $x_{B}$, $z$ and $x_3$ run from zero to unity.

In this work, we use the wave function~\cite{Keum:2000wi,prd65,epjc28-515,li2003,Xiao:2011tx}
\begin{eqnarray}
\Phi_B= \frac{i}{\sqrt{2N_c}} ({ p \hspace{-2.0truemm}/ }_B +m_B) \gamma_5 \phi_B ({\bf k_1}) \;, \label{bmeson}
\end{eqnarray}
for $B^+,B^0$ and $B^0_s$ mesons. And we adopt the widely used distribution
amplitude~\cite{Keum:2000wi,prd65,epjc28-515,li2003,Xiao:2011tx}
\begin{eqnarray}
\phi_B(x,b)&=& N_B x^2(1-x)^2\mathrm{exp} \left  [ -\frac{M_B^2\ x^2}{2 \omega_{B}^2} -\frac{1}{2} (\omega_{B}\; b)^2\right] \;,
 \label{phib}
\end{eqnarray}
for them. With the normalization factor $N_B$ depends on the value of $\omega_B$ and $f_B$,
which is defined through the normalization relation $\int_0^1dx \; \phi_B(x,b=0)=f_B/(2\sqrt{6})$.
$\omega_B = 0.40 \pm0.04$ GeV and $\omega_{B_s}=0.50 \pm 0.05$
GeV~\cite{Keum:2000wi,omega01,omega02} will be employed in the following
numerical calculations.

For the final-state $P$ ($\pi, K, \eta$ or $\eta^\prime$),  we have the wave
functions~\cite{prd65,epjc28-515}
\begin{eqnarray}
\Phi_{P}(P_3,x_3)\equiv \frac{i}{\sqrt{2N_C}}\gamma_5
                    \left [{ p \hspace{-2.0truemm}/ }_3 \phi_{P}^{A}(x_3)+m_{03} \phi_{P}^{P}(x_3)
                    + m_{03} ({ n \hspace{-2.2truemm}/ } { v \hspace{-2.2truemm}/ } - 1)\phi_{P}^{T}(x_3)\right ] \;,
\end{eqnarray}
where $m_{03}$ is the corresponding meson chiral mass, $P_3$ and $x_3$ are the momentum
and the momentum fraction of $P$, respectively.
The expressions of the relevant distribution amplitudes of pion and kaon
mesons are the following~\cite{ball98,ball99,ball9901,ball05,ball06,prd76-074018}:
\begin{eqnarray}
 \phi_{\pi}^A(x) &=& \frac{3f_{\pi}}{\sqrt{6}} x(1-x)[ 1 +0.44C_2^{3/2}(t)] \;, \\
 \phi_{\pi}^P(x) &=& \frac{f_{\pi}}{2\sqrt{6}}[1 +0.43C_2^{1/2}(t)] \;, \\
 \phi_{\pi}^T(x) &=& \frac{f_{\pi}}{2\sqrt{6}}(1-2x)[1+0.55(10x^2-10x+1)]
 \;,\\
 \phi_{K}^A(x) &=& \frac{3f_{K}}{\sqrt{6}}x(1-x)[1+0.17C_1^{3/2}(t)+0.2C_2^{3/2}(t)] \;, \\
 \phi_{K}^P(x) &=& \frac{f_{K}}{2\sqrt{6}} [1+0.24C_2^{1/2}(t)] \;, \\
 \phi_{K}^T(x)
 &=&-\frac{f_{K}}{2\sqrt{6}}[C_1^{1/2} (t)+0.35 C_3^{1/2} (t)] \;.
\end{eqnarray}
The distribution amplitudes $\phi_{\eta_{q(s)}}^{A,P,T}$ ($q$=$u$,$d$) for
$\eta_{q(s)}$ are given as~\cite{ball98,ball99,ball9901,ly14}:
\begin{eqnarray}
 \phi_{\eta_{q(s)}}^A(x) &=&  \frac{f_{q(s)}}{2\sqrt{2N_c} }    6x (1-x)
    \bigg[1+a^{\eta}_1C^{3/2}_1(2x-1)+a^{\eta}_2 C^{3/2}_2(2x-1)+a^{\eta}_4C^{3/2}_4(2x-1)\bigg] \;, \label{piw11}\\
 \phi_{\eta_{q(s)}}^P(x) &=&   \frac{f_{q(s)}}{2\sqrt{2N_c} }
   \bigg[ 1+(30\eta_3-\frac{5}{2}\rho^2_{\eta_{q(s)} } )C^{1/2}_2(2x-1)-3\big[\eta_3\omega_3+\frac{9}{20}\rho^2_{\eta_{q(s)} }
   (1+6a^{\eta }_2)\big] C^{1/2}_4(2x-1)\bigg]  \;,\\
 \phi_{\eta_{q(s)}}^T(x) &=&  \frac{f_{q(s)}}{2\sqrt{2N_c} } (1-2x)   \bigg[ 1+6 (5\eta_3-\frac{1}{2}\eta_3\omega_3
   -\frac{7}{20}\rho^2_{\eta_{q(s)}}-\frac{3}{5}\rho^2_{\eta_{q(s)} }a_2^{\eta} ) (1-10x+10x^2 )\bigg] \;,\quad\quad\label{piw4}
 \end{eqnarray}
with the Gegenbauer moments
\begin{eqnarray}
a^{\eta}_1=0 ,\quad a^{\eta}_2=0.44, \quad a^{\eta}_4=0.25.
\end{eqnarray}
The paprameters $\rho_{\eta_q}=2m_q/m_0^q$ with $m_0^q=1.07 {\rm GeV}$ for $\eta_q$ and
$\rho_{\eta_s}=2m_s/m_0^s$ with $m_0^s=1.92{\rm GeV}$ for $\eta_s$~\cite{prd72-114005}.
The Gegenbauer polynomials $C^{\nu}_n(t)$ ($n=1,2,3,4$ and $\nu=1/2, 3/2$)
above could be found in Ref.~\cite{prd76-074018}.

In this paper, we consider the meson $\eta, \eta^\prime$ as mixtures from $\eta_q$ and $\eta_s$:
\begin{eqnarray}
\left(\begin{array}{c} \eta \\ \eta^{\prime} \end{array} \right)= \left(\begin{array}{cc}
 \cos{\phi} & -\sin{\phi} \\
 \sin{\phi} & \cos{\phi} \\ \end{array} \right)
 \left(\begin{array}{c}
 \eta_q \\ \eta_s \end{array} \right),
\label{eq:e-ep}
\end{eqnarray}
with
\begin{eqnarray}
\eta_q=\frac{1}{\sqrt{2}}\left ( u\bar{u}+d\bar{d}\right ),\quad \eta_s=s\bar{s},
\label{eq:e1-e8}
\end{eqnarray}
The mixtures among the $\eta_q, \eta_s$ and a possible glueball
~\cite{fan:13,glueball-01,glueball-02,PRD87-097501} will
be neglected in this work.
For the decay constant and the mixing angle $\phi$, we have the
forms as~\cite{prd58-114006,prd449-339},
\begin{eqnarray}
f_q=(1.07\pm 0.02)f_{\pi},\quad f_s=(1.34\pm 0.06)f_{\pi},\quad
\phi=39.3^\circ\pm 1.0^\circ, \quad f_{\pi}=0.131~{\rm GeV}.
\end{eqnarray}

The two-pion distribution amplitudes are the same ones as those being used in Ref.~\cite{Wang-2016} ,
\begin{eqnarray}
\Phi_{\pi\pi}^{\rm P}=\frac{1}{\sqrt{2N_c}}[{ p \hspace{-2.0truemm}/ }
\Phi_{v\nu=-}^{I=1}(z,\zeta,\omega^2)+\omega\Phi_{s}^{I=1}(z,\zeta,\omega^2)
+\frac{{p\hspace{-1.5truemm}/}_1{p\hspace{-1.5truemm}/}_2
  -{p\hspace{-1.5truemm}/}_2{p\hspace{-1.5truemm}/}_1}{w(2\zeta-1)}\Phi_{t\nu=+}^{I=1}(z,\zeta,\omega^2)
],
\label{eq:phifunc}
\end{eqnarray}
with
\begin{eqnarray}
\Phi_{v\nu=-}^{I=1}&=&\phi_0=\frac{3F_{\pi}(s)}{\sqrt{2N_c}}z(1-z)\left[1
+a^0_{2\rho}\frac{3}{2}(5(1-2z)^2-1)\right] P_1(2\zeta-1) \;,\\
\Phi_{s}^{I=1}&=&\phi_s=\frac{3F_s(s)}{2\sqrt{2N_c}}(1-2z)\left[1
+a^s_{2\rho}(10z^2-10z+1)\right] P_1(2\zeta-1) \;,\\
\Phi_{t\nu=+}^{I=1}&=&\phi_t=\frac{3F_t(s)}{2\sqrt{2N_c}}(1-2z)^2\left[1
+a^t_{2\rho}\frac{3}{2}(5(1-2z)^2-1)\right] P_1(2\zeta-1) \;,
\end{eqnarray}
where the Legendre polynomial $P_1(2\zeta-1)=2\zeta-1$.
We make tiny corrections of the Gegenbauer moments for the two-pion distribution amplitudes
comparing with those in Ref.~\cite{Wang-2016}. By referring to all the existing data of $B \to P (\rho \to)
\pi\pi$ in Ref.~\cite{pdg2014},
we adjust $a^0_{2\rho}, a^s_{2\rho}, a^t_{2\rho}$ to cater to the data and we have
the new Gegenbauer coefficients $a^0_{2\rho}=0.30, a^s_{2\rho}=0.70, a^t_{2\rho}=-0.40$.

We adopt the same $F_{\pi}(s)$ in this work as that in Ref.~\cite{Wang-2016}, the approximate
relations $F_{s,t}(s)\approx (f_\rho^T/f_\rho) F_\pi(s)$~\cite{Wang-2016} will also be used
in the following section.
By taking the $\rho-\omega$ interference and the excited states into account,
the form factor $F_{\pi}(s)$ can be written in the form of
\beq
F_{\pi}(s)= \left [ {\rm GS}_\rho(s,m_{\rho},\Gamma_{\rho})
\frac{1+c_{\omega} {\rm BW}_{\omega}(s,m_{\omega},\Gamma_{\omega})}{1+c_{\omega}}
+\Sigma c_i {\rm GS}_i(s,m_i,\Gamma_i)\right] \left[ 1+\Sigma c_i\right]^{-1}
\label{eq:fpp}
\eeq
where $s=m^2(\pi\pi)$ is the two-pion invariant mass squared,
$i=(\rho^{\prime}(1450), \rho^{\prime \prime}(1700), \rho^{\prime \prime \prime}(2254))$,
$\Gamma$ is the decay width for the relevant resonance,
$m_{\rho,\omega,i}$ are the masses of the corresponding mesons, respectively.
The function ${\rm GS}_\rho(s,m_{\rho},\Gamma_{\rho})$ has been parameterized as
the Gounaris-Sakurai (GS) model based on the Breit-Wigner (BW) model \cite{GS:68,BW-model}
\begin{equation}
{\rm GS}_\rho(s, m_\rho, \Gamma_\rho) =
\frac{m_\rho^2 [ 1 + d(m_\rho) \Gamma_\rho/m_\rho ] }{m_\rho^2 - s + f(s, m_\rho, \Gamma_\rho)
- i m_\rho \Gamma (s, m_\rho, \Gamma_\rho)}~,
\end{equation}
with the functions
\beq
\Gamma (s, m_\rho, \Gamma_\rho) &=& \Gamma_\rho  \frac{s}{m_\rho^2}
\left( \frac{\beta_\pi (s) }{ \beta_\pi (m_\rho^2) } \right) ^3~,\non
d(m) &=& \frac{3}{\pi} \frac{m_\pi^2}{k^2(m^2)} \ln \left( \frac{m+2 k(m^2)}{2 m_\pi} \right)
   + \frac{m}{2\pi  k(m^2)}
   - \frac{m_\pi^2  m}{\pi k^3(m^2)}~,\non
f(s, m, \Gamma) &=& \frac{\Gamma  m^2}{k^3(m^2)} \left[ k^2(s) [ h(s)-h(m^2) ]
+ (m^2-s) k^2(m^2)  h'(m^2)\right]~,\non
k(s) &=& \frac{1}{2} \sqrt{s}  \beta_\pi (s)~,\non
h(s) &=& \frac{2}{\pi}  \frac{k(s)}{\sqrt{s}}  \ln \left( \frac{\sqrt{s}+2 k(s)}{2 m_\pi} \right).
\eeq
where $\beta_\pi (s) = \sqrt{1 - 4m_\pi^2/s}$. For $\rho(770)$ resonant state, for example,
the measured value
of its resonance width is $\Gamma_\rho = 0.149$ GeV to be used
as input in the numerical calculations.

\section{Numerical results and discussions}\label{sec:3}  

\begin{table}[thp]
\caption{$CP$ averaged branching ratios and direct $CP$-violating asymmetries
of $B_{(s)}\to K (\rho \to) \pi \pi$ decays calculated in PQCD approach together
with experimental data~\cite{pdg2014}}
\label{Presults}
\begin{center}
\begin{tabular}{cccc}
 \hline \hline
{Modes}          &\qquad  & Quasi-two-body results              &  Experiment   \\
\hline
  $B^+ \to K^+(\rho^0\to)\pi^+ \pi^-$             &~~~${\cal B} (10^{-6})$~~~         &$4.04^{+0.75}_{-0.58}(\omega_B)^{+0.24}_{-0.20}(a^t_{2\rho})^{+0.27}_{-0.25}(a^s_{2\rho}) ^{+0.22}_{-0.21}(a^0_{2\rho})$  &$~~3.70\pm{0.50}~~$\\
                                 &$\cal A_{CP} (\%)$    &$50.7^{+3.8}_{-2.6}(\omega_B)^{+3.3}_{-4.7}(a^t_{2\rho})^{+0.0}_{-0.7}(a^s_{2\rho}) ^{+0.9}_{-1.5}(a^0_{2\rho}) $                  &$37.0\pm{10.0}$\\
  $B^0 \to K^+(\rho^-\to)\pi^- \pi^0$            &${\cal B} (10^{-6})$         &$8.17^{+1.93}_{-1.39}(\omega_B)^{+0.36}_{-0.31}(a^t_{2\rho})^{+0.46}_{-0.51}(a^s_{2\rho})\pm{0.43}(a^0_{2\rho})$  &$7.00\pm{0.90}$\\
                                 &$\cal A_{CP} (\%)$    &$39.7^{+2.6}_{-0.6}(\omega_B)^{+5.1}_{-5.4}(a^t_{2\rho})^{+0.5}_{-0.0}(a^s_{2\rho}) ^{+1.0}_{-0.9}(a^0_{2\rho}) $                  &$20.0\pm{11.0}$\\
  $B_s^0 \to K^-(\rho^+\to)\pi^+ \pi^0$           &${\cal B} (10^{-6})$         &$19.68^{+7.63}_{-5.18}(\omega_{B_s})\pm{0.01}(a^t_{2\rho})\pm {0.01}(a^s_{2\rho}) ^{+0.05}_{-0.06}(a^0_{2\rho})$  &$-$\\
                                 &$\cal A_{CP} (\%)$    &$21.8^{+3.7}_{-3.4}(\omega_{B_s})\pm{0.3}(a^t_{2\rho})\pm{0.2}(a^s_{2\rho}) \pm{1.2}(a^0_{2\rho})$                   &$-$\\
  $B^+ \to K^0(\rho^+\to)\pi^+ \pi^0$            &${\cal B} (10^{-6})$         &$8.13^{+1.82}_{-1.23}(\omega_B)\pm{0.87}(a^t_{2\rho})^{+0.44}_{-0.43}(a^s_{2\rho}) ^{+0.36}_{-0.39}(a^0_{2\rho})$  &$8.00\pm{1.50}$\\
                                 &$\cal A_{CP} (\%)$    &$13.8^{+3.1}_{-2.9}(\omega_B)^{+2.2}_{-1.9}(a^t_{2\rho})^{+0.2}_{-0.0}(a^s_{2\rho}) ^{+0.2}_{-0.3}(a^0_{2\rho})$                  &$-12.0\pm{17.0}$\\

  $B^0 \to K^0(\rho^0\to)\pi^+ \pi^-$            &${\cal B} (10^{-6})$         &$4.39^{+1.12}_{-0.81}(\omega_B)\pm0.38(a^t_{2\rho})^{+0.21}_{-0.22}(a^s_{2\rho}) ^{+0.19}_{-0.16}(a^0_{2\rho})$  &$4.70\pm{0.60}$\\
                                 &$\cal A_{CP} (\%)$    &$8.1^{+0.1}_{-0.2}(\omega_B)^{+0.8}_{-0.3}(a^t_{2\rho})^{+0.8}_{-0.6}(a^s_{2\rho})\pm{0.0}(a^0_{2\rho})$                 &$-$\\
  $B_s^0 \to \bar K^0(\rho^0\to)\pi^+ \pi^-$       &${\cal B} (10^{-6})$         &$0.21^{+0.05}_{-0.01}(\omega_{B_s})^{+0.01}_{-0.00}(a^t_{2\rho})^{+0.01}_{-0.00}(a^s_{2\rho}) ^{+0.03}_{-0.01}(a^0_{2\rho})$  &$-$\\
                                 &$\cal A_{CP} (\%)$    &$63.7^{+13.1}_{-15.2}(\omega_{B_s})^{+5.7}_{-7.0}(a^t_{2\rho})^{+3.1}_{-4.0}(a^s_{2\rho}) ^{+1.5}_{-2.0}(a^0_{2\rho}) $                 &$-$\\
 \hline \hline
\end{tabular}
\end{center}
\end{table}

\begin{table}[!hb]
\caption{$CP$ averaged branching ratios and direct $CP$-violating asymmetries
of $B_{(s)}\to \pi (\rho \to) \pi \pi$ decays calculated in PQCD approach together
with experimental data~\cite{pdg2014}}
 \label{PPresults}
\begin{center}
\begin{tabular}{cccc}
 \hline \hline
{Modes}          &\qquad  & Quasi-two-body results              &  Experiment   \\
\hline
  $B^+ \to \pi^+(\rho^0\to)\pi^+ \pi^-$           &~~~${\cal B} (10^{-6})$~~~        &$8.84 ^{+1.48}_{-1.24}(\omega_B)^{+0.12}_{-0.13}(a^t_{2\rho})^{+1.17}_{-1.11}(a^s_{2\rho}) ^{+0.25}_{-0.26}(a^0_{2\rho})$   &$~~8.30\pm{1.20}~~$\\
                                 &$\cal A_{CP} (\%)$    &$-27.5^{+2.3}_{-3.1}(\omega_B)^{+0.9}_{-1.0}(a^t_{2\rho})\pm {1.4}(a^s_{2\rho})\pm{0.9}(a^0_{2\rho}) $                  &$18.0^{+9.0}_{-17.0}$\\
  $B^0 \to \pi^+(\rho^-\to)\pi^- \pi^0$           &${\cal B} (10^{-6})$         &$7.85 ^{+2.60}_{-1.82}(\omega_B)^{+1.77}_{-1.58}(a^t_{2\rho})^{+0.94}_{-0.91}(a^s_{2\rho}) ^{+0.26}_{-0.25}(a^0_{2\rho})$  &$23.00\pm{2.30}$
  \footnote{Branching fraction for the decay $B^0\to\rho^\pm\pi^\mp$ in~\cite{pdg2014}. }\\   
                                 &$\cal A_{CP} (\%)$    &$-31.4^{+3.4}_{-3.3}(\omega_B)^{+3.2}_{-4.0}(a^t_{2\rho})^{+1.1}_{-1.6}(a^s_{2\rho})^{+0.9}_{-0.7}(a^0_{2\rho}) $                 &$-8.0\pm{8.0}$\\
  $B^0 \to \pi^-(\rho^+\to)\pi^+ \pi^0$           &${\cal B} (10^{-6})$         &$18.78 ^{+6.92}_{-4.80}(\omega_B)^{+0.56}_{-0.55}(a^t_{2\rho})^{+0.20}_{-0.21}(a^s_{2\rho}) \pm{0.01}(a^0_{2\rho})$  &$~23.00\pm{2.30}~^a$\\
                                 &$\cal A_{CP} (\%)$    &$8.2^{+1.9}_{-1.5}(\omega_B)\pm {0.3}(a^t_{2\rho})^{+0.2}_{-0.1}(a^s_{2\rho}) ^{+0.6}_{-0.5}(a^0_{2\rho})$ &$13.0\pm{6.0}$\\
  $B_s^0 \to \pi^+(\rho^-\to)\pi^- \pi^0$         &${\cal B} (10^{-6})$         &$0.38\pm{0.05}(\omega_{B_s})\pm0.01(a^t_{2\rho})^{+0.00}_{-0.01}(a^s_{2\rho}) ^{+0.02}_{-0.03}(a^0_{2\rho})$  &$-$\\
                                 &$\cal A_{CP} (\%)$    &$-4.9^{+0.0}_{-1.7}(\omega_{B_s})^{+1.3}_{-4.4}(a^t_{2\rho})^{+0.0}_{-2.5}(a^s_{2\rho}) ^{+0.6}_{-1.5}(a^0_{2\rho})$                  &$-$\\
  $B_s^0 \to \pi^-(\rho^+ \to)\pi^+ \pi^0$         &${\cal B} (10^{-6})$         &$0.41 \pm{0.05}(\omega_{B_s})^{+0.00}_{-0.02}(a^t_{2\rho})\pm{0.01}(a^s_{2\rho}) ^{+0.02}_{-0.03}(a^0_{2\rho})$  &$-$\\
                                 &$\cal A_{CP} (\%)$    &$-36.7^{+0.0}_{-2.5}(\omega_{B_s})^{+2.8}_{-5.4}(a^t_{2\rho})^{+0.1}_{-0.3}(a^s_{2\rho}) ^{+0.0}_{-0.3}(a^0_{2\rho})$ &$-$\\
  $B^+ \to \pi^0(\rho^+\to)\pi^+ \pi^0$           &${\cal B} (10^{-6})$         &$5.53 ^{+2.65}_{-1.79}(\omega_B)^{+0.76}_{-0.71}(a^t_{2\rho})^{+0.49}_{-0.47}(a^s_{2\rho}) ^{+0.00}_{-0.02}(a^0_{2\rho})$  &$10.90\pm{1.40}$\\
                                 &$\cal A_{CP} (\%)$    &$34.9^{+7.3}_{-6.9}(\omega_B)^{+1.6}_{-2.1}(a^t_{2\rho})^{+1.6}_{-1.7}(a^s_{2\rho}) ^{+1.9}_{-1.8}(a^0_{2\rho}) $                  &$2.0\pm{11.0}$\\
  $B^0 \to \pi^0(\rho^0\to)\pi^+ \pi^-$           &${\cal B} (10^{-6})$         &$0.11 ^{+0.06}_{-0.03}(\omega_B)^{+0.02}_{-0.00}(a^t_{2\rho})^{+0.01}_{-0.00}(a^s_{2\rho}) ^{+0.01}_{-0.00}(a^0_{2\rho})$  &$2.00\pm{0.50}$\\
                                 &$\cal A_{CP} (\%)$    &$-14.2^{+17.1}_{-4.3}(\omega_B)^{+3.6}_{-0.7}(a^t_{2\rho})^{+11.3}_{-9.2}(a^s_{2\rho}) ^{+2.8}_{-0.0}(a^0_{2\rho})$ &$-$\\
  $B_s^0 \to \pi^0(\rho^0 \to)\pi^+ \pi^-$         &${\cal B} (10^{-6})$         &$0.35 ^{+0.06}_{-0.05}(\omega_{B_s})\pm{0.01}(a^t_{2\rho})\pm0.00(a^s_{2\rho}) \pm{0.03}(a^0_{2\rho})$  &$-$\\
                                 &$\cal A_{CP}  (\%)$    &$-24.6^{+2.8}_{-0.0}(\omega_{B_s})^{+1.9}_{-0.0}(a^t_{2\rho})^{+0.0}_{-1.6}(a^s_{2\rho}) ^{+0.0}_{-2.6}(a^0_{2\rho})$ &$-$\\
 \hline \hline
\end{tabular}
\end{center}
\end{table}

\begin{table}[thb]

\caption{$CP$ averaged branching ratios and direct $CP$-violating asymmetries
of $B_{(s)}\to \eta^{(\prime)} (\rho \to) \pi \pi$ decays
calculated in PQCD approach together with experimental data~\cite{pdg2014}}
\begin{center}
 \label{PPPresults}

\begin{tabular}{cccc}
 \hline \hline
{Modes}          &\qquad  & Quasi-two-body results              &  Experiment   \\
\hline
  $B^+ \to \eta(\rho^+\to)\pi^+ \pi^0$            &~~~${\cal B} (10^{-6})$~~~        &$6.74^{+2.04}_{-1.50}(\omega_B)^{+0.29}_{-0.27}(a^t_{2\rho})^{+0.10}_{-0.09}(a^s_{2\rho}) ^{+0.02}_{-0.01}(a^0_{2\rho})$  &$~~7.00\pm{2.90}~~$\\
                                 &$\cal A_{CP} (\%)$    &$-0.3^{+0.2}_{-0.0}(\omega_B)^{+0.3}_{-0.2}(a^t_{2\rho})^{+0.0}_{-0.1}(a^s_{2\rho})\pm{0.0}(a^0_{2\rho}) $              &$11.0\pm{11.0}$\\
  $B^0 \to \eta(\rho^0\to)\pi^+ \pi^-$            &${\cal B} (10^{-6})$         &$0.17 ^{+0.03}_{-0.02}(\omega_B)^{+0.03}_{-0.02}(a^t_{2\rho})^{+0.01}_{-0.00}(a^s_{2\rho}) ^{+0.02}_{-0.00}(a^0_{2\rho})$  &$<1.5$\\
                                 &$\cal A_{CP} (\%)$    &$16.3 ^{+3.3}_{-1.6}(\omega_B)^{+9.1}_{-7.2}(a^t_{2\rho})^{+0.0}_{-1.9}(a^s_{2\rho}) ^{+0.0}_{-1.8}(a^0_{2\rho}) $                 &$-$\\
  $B_s^0 \to \eta(\rho^0\to)\pi^+ \pi^-$          &${\cal B} (10^{-6})$         &$0.10 ^{+0.04}_{-0.02}(\omega_{B_s})\pm0.00(a^t_{2\rho})\pm0.00(a^s_{2\rho}) \pm0.00(a^0_{2\rho})$  &$-$\\
                                 &$\cal A_{CP} (\%)$    &$19.2^{+0.1}_{-0.2}(\omega_{B_s})^{+0.0}_{-0.4}(a^t_{2\rho})^{+0.2}_{-0.4}(a^s_{2\rho}) ^{+1.5}_{-1.7}(a^0_{2\rho})$                 &$-$\\
  $B^+ \to \eta^{\prime}(\rho^+\to)\pi^+ \pi^0$   &${\cal B} (10^{-6})$         &$4.56^{+1.44}_{-1.02}(\omega_B)^{+0.16}_{-0.13}(a^t_{2\rho})^{+0.04}_{-0.03}(a^s_{2\rho}) ^{+0.02}_{-0.01}(a^0_{2\rho})$  &$9.70\pm{2.20}$\\
                                 &$\cal A_{CP} (\%)$    &$21.0^{+1.7}_{-1.9}(\omega_B)\pm1.6(a^t_{2\rho})\pm{0.2}(a^s_{2\rho}) ^{+0.3}_{-0.2}(a^0_{2\rho})  $                 &$26.0\pm{17.0}$\\
  $B^0 \to \eta^{\prime}(\rho^0\to)\pi^+ \pi^-$   &${\cal B} (10^{-6})$         &$0.17 ^{+0.05}_{-0.04}(\omega_B)^{+0.01}_{-0.00}(a^t_{2\rho})\pm{0.01}(a^s_{2\rho}) \pm{0.01}(a^0_{2\rho})$ &$<1.3$\\
                                 &$\cal A_{CP} (\%)$    &$12.8^{+0.0}_{-1.2}(\omega_B)^{+21.1}_{-23.6}(a^t_{2\rho})^{+8.3}_{-7.3}(a^s_{2\rho}) ^{+0.1}_{-0.8}(a^0_{2\rho}) $                  &$-$\\
  $B_s^0 \to \eta^{\prime}(\rho^0\to)\pi^+ \pi^-$ &${\cal B} (10^{-6})$         &$0.23 ^{+0.08}_{-0.06}(\omega_{B_s})^{+0.00}_{-0.01}(a^t_{2\rho})\pm0.00(a^s_{2\rho}) ^{+0.00}_{-0.01}(a^0_{2\rho})$  &$-$\\
                                 &$\cal A_{CP} (\%)$    &$37.9^{+0.3}_{-0.5}(\omega_{B_s})\pm0.2(a^t_{2\rho})\pm0.3(a^s_{2\rho}) \pm0.2(a^0_{2\rho})  $                 &$-$\\
 \hline \hline
\end{tabular}
\end{center}
\end{table}

The following input parameters (the masses, decay constants and QCD scale are
in units of  GeV) will be used~\cite{pdg2014} in numerical calculations,
\begin{eqnarray}
\Lambda^{(f=4)}_{ \overline{MS} }&=&0.25, \quad m_{B^0}=5.280, \quad m_{B_s}=5.367,
\quad m_{B^\pm}=5.279, \nonumber\\
m_{\pi^{\pm}}&=&0.140, \quad m_{\pi^0}=0.135, \quad
m_{K^{\pm}}=0.494, \quad m_{K^0}=0.498,\nonumber\\
m_{\eta}&=&0.548, \quad m_{\eta^{\prime}}=0.958, \quad
m_{\rho^0}=0.775, \quad m_{\rho^{\pm}}=0.775,\nonumber\\
m_{b}&=&4.8, \quad m_c=1.275, \quad m_s=0.095, \nonumber\\
f_B&=& 0.19\pm 0.02, \quad f_{B_s}=0.236\pm0.02, \quad \tau_{B^0}=1.519\; ps,\nonumber\\
\tau_{B_{s}}&=&1.512\; ps, \quad \tau_{B^\pm}=1.638\;ps, \quad f_{\rho}=0.216 \pm 0.003, \quad f^T_{\rho}=0.184. \label{eq:inputs}
\end{eqnarray}
The values of the Wolfenstein parameters are the same as given in Ref.~\cite{pdg2014}:
$A=0.814^{+0.023}_{-0.024}, \lambda=0.22537\pm 0.00061$, $\bar{\rho} = 0.117\pm0.021,
\bar{\eta}= 0.353\pm 0.013$.

For the decay $B \to P (\rho \to \pi \pi)$, the differential branching ratio is
written as~\cite{pdg2014},
\begin{eqnarray}
\frac{d{\cal B}}{ds}=\tau_{B}\frac{|\overrightarrow{p_{\pi}}||\overrightarrow{p_P}|}{32\pi^3m^3_{B}}|{\cal A}|^2, \label{expr-br}
\end{eqnarray}
where $\tau_{B}$ is the mean lifetime of $B$ meson, and $s$ is the invariant mass
squared $s=\omega^2=p^2$.
The kinematic variables $|\overrightarrow{p_{\pi}}|$ and $|\overrightarrow{p_P}|$
denote the magnitudes of one $\pi$ meson
in the pion pair and $P$'s momenta in the center-of-mass frame of the pion pair,
\begin{eqnarray}
 |\overrightarrow{p_{\pi}}|=\frac12\sqrt{s-4m^2_{\pi}}, \quad~~
 |\overrightarrow{p_P}|=\frac12  \sqrt{\big[(m^2_{B}-M_3^2)^2-2(m^2_{B}+M_3^2) s
 +s^2 \big]/s}. \label{br-momentum}
\end{eqnarray}

By using the differential branching fraction in Eq.~(\ref{expr-br}) and the decay
amplitudes in the Appendix,
we calculate and list the $CP$ averaged branching rations ($\cal B$)
and direct $CP$-violating asymmetries ($\cal A_{CP}$) for $B_{(s)}\to K (\rho \to \pi \pi) $
in the third column of
Table~\ref{Presults}, $B_{(s)}\to \pi (\rho \to \pi \pi)$ in Table~\ref{PPresults} and
$B_{(s)}\to \eta^{(\prime)} (\rho \to \pi \pi) $ in Table~\ref{PPPresults}.
The first error of these PQCD predictions comes from $\omega_B=(0.40 \pm 0.04)$ {\rm GeV}
for $B^+, B^0$ mesons and
$\omega_{B_s}=(0.50 \pm 0.05)$ {\rm GeV} for $B_s$ meson, the second error is from
$a^t_{2\rho}=-0.40 \pm 0.10$,
while the other two errors result from $a^s_{2\rho}=0.70 \pm 0.20$ and $a^0_{2\rho}
=0.30 \pm 0.05$, respectively.

From the numerical results as shown in above three tables, one can address some issues as follows:
\begin{itemize}
 \item
 Although we have made small changes for the three Gegenbauer moments $a_2^{0,s,t}$,
 the PQCD predictions for
 the branching ratios and direct $CP$ asymmetries of the quasi-two-body decays
 $B^+ \to K^+ (\rho^0 \to)\pi^+\pi^-$, $B^+ \to K^0 (\rho^+ \to)\pi^+\pi^0$,
 $B^0 \to K^+ (\rho^- \to)\pi^-\pi^0$
 and $B^0 \to K^0 (\rho^0 \to)\pi^+\pi^-$ agree well with those as given previously in
Ref.~\cite{Wang-2016}.
The PQCD predictions for the decay rates of these four decay modes are consistent
with currently available
 data~\cite{pdg2014}. For the decay $B^+ \to K^+ (\rho^0 \to)\pi^+\pi^-$,
 the predicted  direct $CP$ asymmetry  ${\cal A_{CP}}=(50.7^{+5.1}_{-5.6}) \%$
 matches the measured
 value $(37.0\pm {10.0}) \% $.

\item
For $B^+ \to \pi^+(\rho^0\to)\pi^+ \pi^-$ decay, the PQCD prediction
for its branching ratio is well consistent
with the world average $(8.3^{+1.2}_{-1.3})\times10^{-6}$ within
errors, but its $CP$ asymmetry is found to
be negative: ${\cal A_{CP}}=(-27.5^{+3.0}_{-3.7}) \%$ numerically.
The {\it BABAR} and LHCb measurements for this quantity, however, prefer a positive $CP$ asymmetry in the
$m(\pi^+\pi^-)$ region peaked at $m_\rho$.
The theoretical predictions based on the QCDF, PQCD and SCET all give
a negative $CP$ asymmetry of order $-0.20$ for $B^+\to \rho^0\pi^+$ (see Table XIII of \cite{CC:Bud}).
This puzzle concerning the sign of ${\cal A_{CP}}(\rho^0\pi^+)$ needs
to be resolved in the near future.

\item
The agreements of PQCD predictions with the data could be achieved
for $B \to \pi(\rho \to) \pi\pi$ decays comparing with
the results in Ref.~\cite{epjc23275}.
The sum of the branching ratios of the $B^0 \to \pi^+(\rho^-\to)\pi^- \pi^0$
and $B^0 \to \pi^-(\rho^+\to)\pi^+ \pi^0$ decays
are in consistent with the world average data.
The calculated ${\cal A_{CP}}(B^0 \to \pi^-(\rho^+\to)\pi^+ \pi^0)
=(8.2^{+2.0}_{-1.6}) \%$ agree with the data $(13.0\pm{6.0}) \%$.
We also obtain  ${\cal A_{CP}}(B^0 \to \pi^+(\rho^-\to)\pi^- \pi^0)
=(-31.4^{+4.9}_{-5.5}) \%$  which needs to be tested
precisely in the future experiments.

\item
We calculated the branching ratios and $CP$ violations of the quasi-two-body
$B \to \eta^{(\prime)}(\rho \to) \pi \pi$ and
find that ${\cal A_{CP}}(B^+ \to \eta(\rho^+\to)\pi^+ \pi^0)=(-0.3^{+0.4}_{-0.2}) \%$ and
${\cal A_{CP}}(B^+ \to \eta^{\prime}(\rho^+\to)\pi^+ \pi^0)=(21.0^{+2.4}_{-2.5}) \%$
agree with the data.
The contributions of the tree diagrams are larger than the penguin ones by roughly a
factor of 200 for the decay
  $B^+ \to \eta(\rho^+\to)\pi^+ \pi^0$ and a factor of 40 for the $B^+
  \to \eta^{\prime}(\rho^+\to)\pi^+ \pi^0$.
The tree contribution is therefore dominant for the decay $B^+ \to \eta(\rho^+\to)\pi^+ \pi^0$.
Its direct $CP$ asymmetry  is really small in size.
We also give predictions for $B^0 \to \eta(\rho^0\to)\pi^+ \pi^-$ and
$B^0 \to \eta^{\prime}(\rho^0\to)\pi^+ \pi^-$ decays.

\item
For all the $B_s \to K(\pi,\eta^{(\prime)})\rho \to K(\pi,\eta^{(\prime)}) \pi\pi$
decay channels considered
in this paper, we can compare our PQCD predictions with those as given in the Table
VII and Table VIII of
Refs.~\cite{prd76-074018,prd80-114026}.
From the $CP$ averaged branching ratios, for example,
our results for decays $B_s \to K(\pi,\eta^{(\prime)})\rho \to K(\pi,\eta^{(\prime)}) \pi\pi$  are
a little larger than the corresponding ones in Table VII of Ref.~\cite{prd76-074018}.
As verified in Ref.~\cite{Wang-2016}, it may be more appropriate to treat
$B \to K(\pi,\eta^{(\prime)})\rho$ as the
quasi-two-body decays.
For $B_s^0 \to \pi^-(\rho^+ \to)\pi^+ \pi^0$ and $B_s^0 \to \eta(\rho^0\to)\pi^+ \pi^-$ decays,
we obtain sizeable negative $CP$ asymmetries which could be examined in the
forthcoming experiments.
Our PQCD predictions for  the direct $CP$ asymmetries of  $B_s^0 \to K^-(\rho^+\to)\pi^+ \pi^0$,
$B_s^0 \to \bar K^0(\rho^0\to)\pi^+ \pi^-$  $B_s^0 \to \eta(\rho^0\to)\pi^+ \pi^-$ and
$B_s^0 \to \eta^{\prime}(\rho^0\to)\pi^+ \pi^-$ decays are positive and sizable.

\item
  For the $B^0 \to \pi^0 \rho^0 \to \pi^0\pi^+ \pi^-$ decay process, PQCD prediction
  is ${\cal B}=(0.11^{+0.07}_{-0.03}) \times 10^{-6}$ at leading-order in
  the quasi-two-body framework
  in this work, such a branching ratio is much smaller than the value
  $(2.0\pm 0.5)\times 10^{-6}$ in~\cite{pdg2014}. Similar with the
  $\pi\pi, \pi K$ or $\rho\rho$ puzzles discussed in
  Refs.~\cite{prd72-074007,prd73-015003,jhep09-038,jhep05-056,ijmp26-1273,prd72-114005,prd90-014029,
  prd91-114019,prd93-014024}, the $B \to \pi\rho$ puzzle has been noticed
  by some groups~\cite{prd62-036001,
  prl86-216,prd65-094004,prd66-034019,epja18-543,hep0306294,hep0306298}.
For example, in Ref.~\cite{prd65-094004}, the authors examined the role
of $\sigma \pi$ channel in the Dalitz plot
analysis of $\rho \pi$ decays and concluded that the effect of $\sigma$
to $B^0 \to \rho^0\pi^0$ is not important.
  While, in~\cite{prd66-034019}, the authors found that $B^0 \to\rho^0 \pi^0$ process
  could receive large contributions from
  the heavy-meson $B^*$ and $B_0$ backgrounds.
  Since the isospin-violating effect is visible in the $e^+e^- \to \pi^+ \pi^-$ data at
  $s=m^2_{\omega}$~\cite{prd86-032013}, the $\rho^0$-$\omega$ mixing need to be taken into
  studies~\cite{prd65-094004,plb386-413,prl80-1834,epjc31-215,jpg31-199, prd71-074017}.
  We leave the gap between the data in~\cite{pdg2014} and the PQCD prediction ${\cal B}=(0.11^{+0.07}_{-0.03}) \times 10^{-6}$
  to the future studies.

\end{itemize}

\begin{table}[tb]
\caption{For the measured decay mode $B^+ \to K^+(\rho^0\to)\pi^+ \pi^-$,
the $\Gamma_\rho$-dependence of the PQCD predictions for the branching ratios
and the direct $CP$-violating asymmetries, assuming $0\leq \Gamma\rho \leq 0.149$ GeV. }
\begin{center} \label{width}
\begin{tabular}{c|c|c|c|c|c|c|c}
\hline \hline
$\Gamma_\rho $(GeV) &$0$ &$0.005$ &$0.015$ &$0.060$ &$0.090$ &$0.120$ &$0.149$  \\ \hline
${\cal B} (10^{-6})$ &$5370.2$ &$105.5$ &$35.4$ &$9.2$ &$6.3$ &$4.9$ &$4.0$ \\ \hline
$\cal A_{CP} (\%)$   &$50.9$ &$53.3$ &$52.8$ &$51.8$ &$51.2$ &$50.8$ &$50.7$  \\ \hline \hline
\end{tabular} \end{center} \end{table}

For the considered $B/B_s \to P(\rho\to) \pi\pi$ decays, we know that the
introduction of the resonance width $\Gamma_\rho$ is one of the crucial
differences between the two-body formalism and the quasi-two-body one and
may play an important role in our theoretical predictions
for the $CP$ averaged branching ratios and the $CP$-violating asymmetries.
In order to check the $\Gamma_\rho$-dependence of these physical
observables,
we vary $\Gamma_\rho$ in Eqs.~(26-27) in the range of $0 \leq \Gamma_\rho \leq 0.149$ GeV
and list our PQCD predictions in Table~\ref{width}.
For the sake of simplicity, we take the experimentally  measured decay mode
$B^+ \to K^+(\rho^0\to)\pi^+ \pi^-$ as an example, and
make numerical calculations for the seven fixed values of $\Gamma_\rho$.
From the numerical results in Table~\ref{width}, we find easily that
\begin{itemize}
\item
Our PQCD predictions for the branching ratios are very sensitive on the variations of
the given value of the resonance width $\Gamma_\rho$.
For $\Gamma_\rho=\Gamma_\rho^{exp}=0.149$ GeV,
the PQCD prediction ${\cal B}(B^+ \to K^+(\rho^0\to)\pi^+ \pi^-)\approx 4.0\times 10^{-6}$
agrees well with the measured value $(3.7\pm 0.5)\times 10^{-6}$ \cite{pdg2014}.

\item
For CP asymmetries ${\cal A}_{CP}$, the $\Gamma_\rho$-dependence is indeed negligible.

\end{itemize}

\section{CONCLUSION}

In this paper, we calculated the $CP$-averaged branching ratios and direct $CP$-violating
asymmetries of
the quasi-two-body decays $B_{(s)} \to (\pi,K,\eta,\eta^\prime ) \rho
\to (\pi,K,\eta,\eta^\prime ) \pi\pi$
by using the PQCD factorization approach.
The two-pion distribution amplitude $\Phi_{\pi\pi}^{\rm P}$ with the $P$-wave timelike form
factor $F_\pi$ was employed to describe the resonant state $\rho$ and
its interactions with the pion pair.
General agreements between the PQCD predictions and the data achieved
by making a little adjustments
of the Gegenbauer moments of the $P$-wave two-pion distribution amplitudes.
We listed the PQCD predictions for those considered decay channels,
which will be tested at the LHCb and Belle-II experiment.

From the numerical results, we found the following points:
\begin{itemize}
\item
Except for the $B \to \pi^0 \rho^0 \to \pi^0 (\pi^+ \pi^-)$ decay mode, the PQCD predictions
for the branching ratios of other $B_{(s)} \to  (\pi,K,\eta,\eta^\prime ) \rho
\to (\pi,K,\eta,\eta^\prime ) \pi\pi$
decays agree with currently available data within errors.

\item
For ${\cal B}(B \to \pi^0 \rho^0 \to \pi^0(\pi^+ \pi^-))$ decay, the PQCD prediction is about
$(0.11^{+0.07}_{-0.03}) \times 10^{-6}$  and is much smaller than the measured
one: $(2.0\pm 0.5)\times 10^{-6}$.

\item
For $B^+ \to \pi^+(\rho^0\to)\pi^+ \pi^-$ decay mode, we found a negative $CP$ asymmetry
$(-27.5^{+3.0}_{-3.7}) \%$, which agrees with theoretical predictions based
on QCDF or other factorization approaches, but different in sign from
the measured ones in the $m(\pi^+\pi^-)$ region peaked at $m_\rho$, as reported
by {\it BABAR} and LHCb Collaboration.
Such difference should be tested in the forthcoming experimental measurements.

\end{itemize}

\begin{acknowledgments}
Many thanks to Hsiang-nan Li, Cai-Dian L\"u, Xin Liu, Rui Zhou and Wei Wang for valuable discussions.
This work was supported by the National Natural Science Foundation of
China under the No.~11235005 and No.~11547038.

\end{acknowledgments}
\appendix

\section{Decay amplitudes}

The total decay amplitude for each considered decay mode in this work are given as follows:

\begin{eqnarray}
{\cal A}(B^+ \to K^+(\rho^0 \to)\pi^+ \pi^-) &=& \frac{G_F} {2} \big\{V_{ub}^*V_{us}[(\frac{C_1}{3}+C_2)(F^{LL}_{e\rho}+F^{LL}_{a\rho})+(C_1+\frac{C_2}{3})F^{LL}_{eP}+C_2M^{LL}_{eP}\nonumber\\
&+&C_1(M^{LL}_{e\rho}+M^{LL}_{a\rho})]-V_{tb}^*V_{ts}[(\frac{C_3}{3}+C_4+\frac{C_9}{3}+C_{10})(F^{LL}_{e\rho}+F^{LL}_{a\rho})\nonumber\\
&+&(\frac{C_5}{3}+C_6+\frac{C_7}{3}+C_8)(F^{SP}_{e\rho}+F^{SP}_{a\rho})+(C_3+C_9)(M^{LL}_{e\rho}+M^{LL}_{a\rho})\nonumber\\
&+&(C_5+C_7)(M^{LR}_{e\rho}+M^{LR}_{a\rho})+\frac{3 C_8}{2} M^{SP}_{eP}+\frac{3 C_{10} }{2} M^{LL}_{eP} \nonumber\\
&+&\frac{3}{2}(C_7+\frac{C_8}{3}+C_9+\frac{C_{10}}{3})F^{LL}_{eP}]\big\} \;, \label{amp1}
\end{eqnarray}
 \begin{eqnarray}
{\cal A}(B^0 \to K^+(\rho^- \to)\pi^-\pi^0) &=& \frac{G_F} {\sqrt{2}} \big\{V_{ub}^*V_{us}[(\frac{C_1}{3}+C_2)F^{LL}_{e\rho}+C_1 M^{LL}_{e\rho}]-V_{tb}^*V_{ts}[(C_3+C_9)M^{LL}_{e\rho}\nonumber\\
&+&(\frac{C_3}{3}+C_4+\frac{C_9}{3}+C_{10})F^{LL}_{e\rho}+(\frac{C_5}{3}+C_6+\frac{C_7}{3}+C_8)F^{SP}_{e\rho}\nonumber\\
&+&(C_5+C_7)M^{LR}_{e\rho}+(\frac{C_3}{3}+C_4-\frac{1}{2}(\frac{C_9}{3}+C_{10}))F^{LL}_{a\rho}+(C_3-\frac{C_9}{2})M^{LL}_{a\rho}\nonumber\\ &+&(\frac{C_5}{3}+C_6-\frac{1}{2}(\frac{C_7}{3}+C_8))F^{SP}_{a\rho}+(C_5-\frac{C_7}{2})M^{LR}_{a\rho}]\big\} \;,\label{amp2}
\end{eqnarray}
\begin{eqnarray}
 {\cal A}(B_s^0 \to K^-(\rho^+\to)\pi^+ \pi^0) &=&  \frac{G_F} {\sqrt{2}} \big\{V_{ub}^*V_{ud}[(\frac{C_1}{3}+C_2)F^{LL}_{eP}+C_1 M^{LL}_{eP}]-V_{tb}^*V_{td}[(C_3+C_9)M^{LL}_{eP}\nonumber\\
&+&(\frac{C_3}{3}+C_4+\frac{C_9}{3}+C_{10})F^{LL}_{eP}+(C_5+C_7)M^{LR}_{eP}\nonumber\\
&+&(\frac{C_3}{3}+C_4-\frac{1}{2}(\frac{C_9}{3}+C_{10}))F^{LL}_{aP}+(\frac{C_5}{3}+C_6-\frac{1}{2}(\frac{C_7}{3}+C_8))F^{SP}_{aP}\nonumber\\
&+&(C_3-\frac{C_9}{2})M^{LL}_{aP}+(C_5-\frac{C_7}{2})M^{LR}_{aP}]\big\} \;,\label{amp3}
\end{eqnarray}
\begin{eqnarray}
 {\cal A}(B^+ \to K^0(\rho^+\to)\pi^+ \pi^0) &=& \frac{G_F} {\sqrt{2}}\big\{V_{ub}^*V_{us}[(\frac{C_1}{3}+C_2)F^{LL}_{a\rho}+C_1 M^{LL}_{a\rho}]-V_{tb}^*V_{ts}[(C_3-\frac{C_9}{2})M^{LL}_{e\rho}\nonumber\\
&+&(\frac{C_3}{3}+C_4-\frac{1}{2}(\frac{C_9}{3}+C_{10}))F^{LL}_{e\rho}+(\frac{C_5}{3}+C_6-\frac{1}{2}(\frac{C_7}{3}+C_8))F^{SP}_{e\rho}\nonumber\\
&+&(C_5-\frac{C_7}{2})M^{LR}_{e\rho}+(\frac{C_3}{3}+C_4+\frac{C_9}{3}+C_{10})F^{LL}_{a\rho}+(C_3+C_9)M^{LL}_{a\rho}\nonumber\\
&+&(\frac{C_5}{3}+C_6+\frac{C_7}{3}+C_8)F^{SP}_{a\rho}+(C_5+C_7)M^{LR}_{a\rho}]\big\} \;,\label{amp4}
\end{eqnarray}
\begin{eqnarray}
{\cal A}(B^0 \to K^0(\rho^0\to)\pi^+ \pi^-) &=& \frac{G_F} {2}\big\{V_{ub}^*V_{us}[(C_1+\frac{C_2}{3})F^{LL}_{eP}+C_2 M^{LL}_{eP}]-V_{tb}^*V_{ts}[\frac{3 C_8}{2}M^{SP}_{eP}\nonumber\\
&-&(\frac{C_3}{3}+C_4-\frac{1}{2}(\frac{C_9}{3}+C_{10}))(F^{LL}_{e\rho}+F^{LL}_{a\rho})-(C_3-\frac{C_9}{2})(M^{LL}_{e\rho}+M^{LL}_{a\rho})\nonumber\\
&-&(\frac{C_5}{3}+C_6-\frac{1}{2}(\frac{C_7}{3}+C_8))(F^{SP}_{e\rho}+F^{SP}_{a\rho})-(C_5-\frac{C_7}{2})(M^{LR}_{e\rho}+M^{LR}_{a\rho})\nonumber\\
&+&\frac{3}{2}(C_7+\frac{C_8}{3}+C_9+\frac{C_{10}}{3})F^{LL}_{eP}+\frac{3 C_{10}}{2}M^{LL}_{eP}]\big\} \;,\label{amp5}
 \end{eqnarray}
 \begin{eqnarray}
{\cal A}(B_s^0 \to K^0(\rho^0\to)\pi^+ \pi^-) &=& \frac{G_F} {2}\big\{V_{ub}^*V_{ud}[(C_1+\frac{C_2}{3})F^{LL}_{eP}+C_2 M^{LL}_{eP}]-V_{tb}^*V_{td}[\frac{3 C_8}{2}M^{SP}_{eP}\nonumber\\
&+&(-\frac{C_3}{3}-C_4+\frac{5 C_9}{3}+C_{10}+\frac{3}{2}(C_7+\frac{C_8}{3}))F^{LL}_{eP}\nonumber\\
&+&(-C_3+\frac{C_9}{2}+\frac{3 C_{10}}{2})M^{LL}_{eP}-(C_5-\frac{C_7}{2})(M^{LR}_{eP}+M^{LR}_{aP})\nonumber\\
&-&(\frac{C_3}{3}+C_4-\frac{1}{2}(\frac{C_9}{3}+C_{10}))F^{LL}_{aP})-(C_3-\frac{C_9}{2})M^{LL}_{aP}\nonumber\\
&-&(\frac{C_5}{3}+C_6-\frac{1}{2}(\frac{C_7}{3}+C_8))F^{SP}_{aP}]\big\} \;,\label{amp6}
\end{eqnarray}
 \begin{eqnarray}
{\cal A}(B^+ \to \pi^+(\rho^0\to)\pi^+ \pi^-) &=& \frac{G_F} {2}\big\{V_{ub}^*V_{ud}[(\frac{C_1}{3}+C_2)(F^{LL}_{e\rho}+F^{LL}_{a\rho}-F^{LL}_{aP})+(C_1+\frac{C_2}{3})F^{LL}_{eP}\nonumber\\
&+&C_1 (M^{LL}_{e\rho}+M^{LL}_{a\rho}-M^{LL}_{aP})+C_2 M^{LL}_{eP}]-V_{tb}^*V_{td}[\frac{3 C_8}{2} M^{SP}_{eP}\nonumber\\
&+&(\frac{C_3}{3}+C_4+\frac{C_9}{3}+C_{10})(F^{LL}_{e\rho}+F^{LL}_{a\rho}-F^{LL}_{aP})\nonumber\\
&+&(C_3+C_9)(M^{LL}_{e\rho}+M^{LL}_{a\rho}-M^{LL}_{aP})+(-C_5+\frac{C_7}{2})M^{LR}_{eP}\nonumber\\
&+&(\frac{C_5}{3}+C_6+\frac{C_7}{3}+C_8)(F^{SP}_{e\rho}+F^{SP}_{a\rho}-F^{SP}_{aP})\nonumber\\
&+&(C_5+C_7)(M^{LR}_{e\rho}+M^{LR}_{a\rho}-M^{LR}_{aP})+(-\frac{C_3}{3}-C_4+\frac{5}{3}C_9\nonumber\\
&+&C_{10}+\frac{3}{2}(C_7+\frac{C_8}{3}))F^{LL}_{eP}+(-C_3+\frac{C_9}{2}+\frac{3 C_{10}}{2})M^{LL}_{eP}]\big\} \;,\label{amp7}
\end{eqnarray}
 \begin{eqnarray}
 {\cal A}(B^0 \to \pi^-(\rho^+\to)\pi^+ \pi^0) &=& \frac{G_F} {\sqrt{2}}
 \big\{V_{ub}^*V_{ud}[(C_1+\frac{C_2}{3})F^{LL}_{a\rho}+(\frac{C_1}{3}+C_2)F^{LL}_{eP}+C_2 M^{LL}_{a\rho}+C_1 M^{LL}_{eP}]\nonumber\\
&-&V_{tb}^*V_{td}[(\frac{C_3}{3}+C_4+\frac{C_9}{3}+C_{10})F^{LL}_{eP}+(C_4+C_{10})M^{LL}_{a\rho}\nonumber\\
&+&(C_3+\frac{C_4}{3}-C_5-\frac{C_6}{3}-C_7-\frac{C_8}{3}+C_9+\frac{C_{10}}{3})F^{LL}_{a\rho}\nonumber\\
&+&(C_3+C_9)M^{LL}_{eP}+(C_5+C_7)M^{LR}_{eP}+(C_5-\frac{C_7}{2})M^{LR}_{aP}\nonumber\\
&+&(\frac{4}{3}(C_3+C_4-\frac{C_9}{2}-\frac{C_{10}}{2})-C_5-\frac{C_6}{3}+\frac{1}{2}(C_7+\frac{C_8}{3}))F^{LL}_{aP}\nonumber\\
&+&(\frac{C_5}{3}+C_6-\frac{1}{2}(\frac{C_7}{3}+C_8))F^{SP}_{aP}+(C_6-\frac{C_8}{2})M^{SP}_{aP}\nonumber\\
&+&(C_3+C_4-\frac{C_9}{2}-\frac{C_{10}}{2})M^{LL}_{aP}+(C_6+C_8)M^{SP}_{a\rho}]\big\} \;,\label{amp8}
\end{eqnarray}
 \begin{eqnarray}
 {\cal A}(B^0 \to \pi^+(\rho^-\to)\pi^- \pi^0) &=& \frac{G_F} {\sqrt{2}}
 \big\{V_{ub}^*V_{ud}[(\frac{C_1}{3}+C_2)F^{LL}_{e\rho}+(C_1+\frac{C_2}{3})F^{LL}_{aP}+C_1 M^{LL}_{e\rho}+C_2 M^{LL}_{aP}]\nonumber\\
 &-&V_{tb}^*V_{td}[(\frac{C_3}{3}+C_4+\frac{C_9}{3}+C_{10})F^{LL}_{e\rho}+(C_3+C_9)M^{LL}_{e\rho}\nonumber\\
&+&(\frac{C_5}{3}+C_6+\frac{C_7}{3}+C_8)F^{SP}_{e\rho}+(C_5+C_7)M^{LR}_{e\rho}+(C_6+C_8)M^{SP}_{aP}\nonumber\\
&+&(\frac{4}{3}(C_3+C_4-\frac{C_9}{2}-\frac{C_{10}}{2})-C_5-\frac{C_6}{3}+\frac{1}{2}(C_7+\frac{C_8}{3}))F^{LL}_{a\rho}\nonumber\\
&+&(\frac{C_5}{3}+C_6-\frac{1}{2}(\frac{C_7}{3}+C_8))F^{SP}_{a\rho}+(C_3+C_4-\frac{C_9}{2}-\frac{C_{10}}{2})M^{LL}_{a\rho}\nonumber\\
&+&(C_5-\frac{C_7}{2})M^{LR}_{a\rho}+(C_6-\frac{C_8}{2})M^{SP}_{a\rho}+(C_4+C_{10})M^{LL}_{aP}\nonumber\\
&+&(C_3+\frac{C_4}{3}-C_5-\frac{C_6}{3}-C_7-\frac{C_8}{3}+C_9+\frac{C_{10}}{3})F^{LL}_{aP}]\big\} \;,\label{amp9}
\end{eqnarray}
 \begin{eqnarray}
 {\cal A}(B_s^0 \to \pi^-(\rho^+\to)\pi^+ \pi^0) &=& \frac{G_F} {\sqrt{2}}
\big\{V_{ub}^*V_{us}[(C_1+\frac{C_2}{3})F^{LL}_{a\rho}+C_2 M^{LL}_{a\rho}]-V_{tb}^*V_{ts}[(C_6+C_8)M^{SP}_{a\rho}\nonumber\\
&+&(C_3+\frac{C_4}{3}-C_5-\frac{C_6}{3}-C_7-\frac{C_8}{3}+C_9+\frac{C_{10}}{3})F^{LL}_{a\rho}\nonumber\\
&+&(C_3+\frac{C_4}{3}-\frac{1}{2}(C_9+\frac{C_{10}}{3})-C_5-\frac{C_6}{3}+\frac{1}{2}(C_7+\frac{C_8}{3}))F^{LL}_{aP}\nonumber\\
&+&(C_4-\frac{C_{10}}{2})M^{LL}_{aP}+(C_6-\frac{C_8}{2})M^{SP}_{aP}+(C_4+C_{10})M^{LL}_{a\rho}]\big\} \;,\label{amp10}
\end{eqnarray}
\begin{eqnarray}
 {\cal A}(B_s^0 \to \pi^+(\rho^-\to)\pi^- \pi^0) &=& \frac{G_F} {\sqrt{2}}
 \big\{V_{ub}^*V_{us}[(C_1+\frac{C_2}{3})F^{LL}_{aP}+C_2 M^{LL}_{aP}]-V_{tb}^*V_{ts}[(C_4-\frac{C_{10}}{2})M^{LL}_{a\rho}
 \nonumber\\
&+&(C_3+\frac{C_4}{3}-\frac{1}{2}(C_9+\frac{C_{10}}{3})-C_5-\frac{C_6}{3}+\frac{1}{2}(C_7+\frac{C_8}{3}))F^{LL}_{a\rho}\nonumber\\
&+&(C_6-\frac{C_8}{2})M^{SP}_{a\rho}+(C_4+C_{10})M^{LL}_{aP}+(C_6+C_8)M^{SP}_{aP}\nonumber\\
&+&(C_3+\frac{C_4}{3}-C_5-\frac{C_6}{3}-C_7-\frac{C_8}{3}+C_9+\frac{C_{10}}{3})F^{LL}_{aP}]\big\} \;,\label{amp11}
\end{eqnarray}
\begin{eqnarray}
{\cal A}(B^+ \to \pi^0(\rho^+\to)\pi^+ \pi^0) &=& \frac{G_F} {2}
\big\{V_{ub}^*V_{ud}[(C_1+\frac{C_2}{3})F^{LL}_{e\rho}+(\frac{C_1}{3}+C_2)(-F^{LL}_{a\rho}+F^{LL}_{eP}+F^{LL}_{aP})\nonumber\\
&+&C_2 M^{LL}_{e\rho}+C_1(- M^{LL}_{a\rho}+M^{LL}_{eP}+M^{LL}_{aP})]-V_{tb}^*V_{td}[\frac{3 C_8}{2}M^{SP}_{e\rho}\nonumber\\
&+&(-\frac{C_3}{3}-C_4-\frac{3}{2}(C_7+\frac{C_8}{3})+\frac{5 C_9}{3}+C_{10})F^{LL}_{e\rho}\nonumber\\
&+&(-\frac{C_5}{3}-C_6+\frac{1}{2}(\frac{C_7}{3}+C_8))F^{SP}_{e\rho}+(-C_3+\frac{C_9}{2}+\frac{3 C_{10}}{2})M^{LL}_{e\rho}
\nonumber\\
&+&(\frac{C_3}{3}+C_4+\frac{C_9}{3}+C_{10})(-F^{LL}_{a\rho}+F^{LL}_{eP}+F^{LL}_{aP})\nonumber\\
&+&(\frac{C_5}{3}+C_6+\frac{C_7}{3}+C_8)(-F^{SP}_{a\rho}+F^{SP}_{aP})+(-C_5+\frac{C_7}{2})M^{LR}_{e\rho}\nonumber\\
&+&(C_3+C_9)(-M^{LL}_{a\rho}+M^{LL}_{eP}+M^{LL}_{aP})\nonumber\\
&+&(C_5+C_7)(-M^{LR}_{a\rho}+M^{LR}_{eP}+M^{LR}_{aP})]\big\} \;,\label{amp12}
\end{eqnarray}
\begin{eqnarray}
 {\cal A}(B^0 \to \pi^0(\rho^0\to)\pi^+ \pi^-) &=& -\frac{G_F} {2\sqrt{2}}
 \big\{V_{ub}^*V_{ud}[(C_1+\frac{C_2}{3})(F^{LL}_{e\rho}-F^{LL}_{a\rho}+F^{LL}_{eP}-F^{LL}_{aP})\nonumber\\
&+&C_2(M^{LL}_{e\rho}-M^{LL}_{a\rho}+M^{LL}_{eP}-M^{LL}_{aP})]-V_{tb}^*V_{td}[\frac{3 C_8}{2}(M^{SP}_{e\rho}+M^{SP}_{eP})\nonumber\\
&+&(-\frac{C_3}{3}-C_4-\frac{3}{2}(C_7+\frac{C_8}{3})+\frac{5 C_9}{3}+C_{10})(F^{LL}_{e\rho}+F^{LL}_{eP}) \nonumber\\
&+&(-\frac{C_5}{3}-C_6+\frac{1}{2}(\frac{C_7}{3}+C_8))F^{SP}_{e\rho}+(-C_3+\frac{C_9}{2}+\frac{3 C_{10}}{2})(M^{LL}_{e\rho}+M^{LL}_{eP})\nonumber\\
&+&(-C_5+\frac{C_7}{2})(M^{LR}_{e\rho}+M^{LL}_{eP})-(2C_6+\frac{C_8}{2})(M^{SP}_{a\rho}+M^{SP}_{sP})\nonumber\\
&-&(\frac{7 C_3}{3}+\frac{5 C_4}{3}-2(C_5+\frac{C_6}{3})-\frac{1}{2}(C_7+\frac{C_8}{3}-\frac{2}{3}(C_9-C_{10})))(F^{LL}_{a\rho}+F^{LL}_{aP})\nonumber\\
&-&(\frac{C_5}{3}+C_6-\frac{1}{2}(\frac{C_7}{3}+C_8))(F^{SP}_{a\rho}+F^{SP}_{aP})-(C_5-\frac{C_7}{2})(M^{LR}_{a\rho}+M^{LR}_{aP})\nonumber\\
&-&(C_3+2C_4-\frac{C_9}{2}+\frac{C_{10}}{2})(M^{LL}_{a\rho}+M^{LL}_{aP})]\big\} \;,\label{amp13}
\end{eqnarray}
 \begin{eqnarray}
{\cal A}(B_s^0 \to \pi^0(\rho^0\to)\pi^+ \pi^-) &=& \frac{G_F} {2\sqrt{2}}
\big\{V_{ub}^*V_{us}[(C_1+\frac{C_2}{3})(F^{LL}_{a\rho}+F^{LL}_{aP})+C_2(M^{LL}_{a\rho}+M^{LL}_{aP})]\nonumber\\
&-&V_{tb}^*V_{ts}[(2(C_3+\frac{C_4}{3}-C_5-\frac{C_6}{3})-\frac{1}{2}(C_7+\frac{C_8}{3}-C_9-\frac{C_{10}}{3}))(F^{LL}_{a\rho}+F^{LL}_{aP})\nonumber\\
&+&(2C_4+\frac{C_{10}}{2})(M^{LL}_{a\rho}+M^{LL}_{aP})+(2C_6+\frac{C_8}{2})(M^{SP}_{a\rho}+M^{SP}_{aP})]\big\} \;,\label{amp14}
\end{eqnarray}
 \begin{eqnarray}
{\cal A}(B^+ \to \eta_q(\rho^+\to)\pi^+ \pi^0) &=& \frac{G_F} {2}
\big\{V_{ub}^*V_{ud}[(C_1+\frac{C_2}{3})F^{LL}_{e\rho}+(\frac{C_1}{3}+C_2)(F^{LL}_{a\rho}+F^{LL}_{eP}+F^{LL}_{aP})\nonumber\\
&+&C_2 M^{LL}_{e\rho}+C_1(M^{LL}_{a\rho}+M^{LL}_{eP}+M^{LL}_{aP})]-V_{tb}^*V_{td}[(C_5-\frac{C_7}{2})M^{LR}_{e\rho}\nonumber\\
&+&(\frac{7 C_3}{3}+\frac{5 C_4}{3}-2(C_5+\frac{C_6}{3})-\frac{1}{2}(C_7+\frac{C_8}{3}-\frac{2}{3}(C_9-C_{10})))F^{LL}_{e\rho}\nonumber\\
&+&(\frac{C_5}{3}+C_6-\frac{1}{2}(\frac{C_7}{3}+C_8))F^{SP}_{e\rho}+(C_3+2C_4-\frac{C_9}{2}+\frac{C_{10}}{2})M^{LL}_{e\rho}\nonumber\\
&+&(\frac{C_3}{3}+C_4+\frac{C_9}{3}+C_{10})(F^{LL}_{a\rho}+F^{LL}_{eP}+F^{LL}_{aP})\nonumber\\
&+&(\frac{C_5}{3}+C_6+\frac{C_7}{3}+C_8)(F^{SP}_{a\rho}+F^{SP}_{aP})+(C_3+C_9)(M^{LL}_{a\rho}+M^{LL}_{eP}+M^{LL}_{aP})\nonumber\\
&+&(C_5+C_7)(M^{LR}_{a\rho}+M^{LR}_{eP}+M^{LR}_{aP})+(2C_6+\frac{C_8}{2})M^{SP}_{e\rho}]\big\} \;,\label{amp15}
\end{eqnarray}
 \begin{eqnarray}
 {\cal A}(B^+ \to \eta_s(\rho^+\to)\pi^+ \pi^0) &=& \frac{G_F} {\sqrt{2}}
 \big\{-V_{tb}^*V_{td}[(C_3+\frac{C_4}{3}-C_5-\frac{C_6}{3}+\frac{1}{2}(C_7+\frac{C_8}{3}-C_9-\frac{C_{10}}{3}))F^{LL}_{e\rho}\nonumber\\
&+&(C_4-\frac{C_{10}}{2})M^{LL}_{e\rho}+(C_6-\frac{C_8}{2})M^{SP}_{e\rho}]\big\} \;,\label{amp16}
\end{eqnarray}
\begin{eqnarray}
{\cal A}(B^+ \to \eta( \rho^+\to)\pi^+ \pi^0) &=& {\cal A}(B^+ \to \rho^+ \eta_q) \cos{\phi}-{\cal A}(B^+ \to \rho^+ \eta_s)\sin{\phi} \;,\label{amp17}\\
{\cal A}(B^+ \to \eta^{\prime}(\rho^+\to)\pi^+ \pi^0) &=&  {\cal A}(B^+ \to \rho^+ \eta_q)\sin{\phi}+ {\cal A}(B^+ \to \rho^+ \eta_s) \cos{\phi} \;,\label{amp18}
\end{eqnarray}
\begin{eqnarray}
{\cal A}(B^0 \to \eta_q(\rho^0\to)\pi^+ \pi^-) &=& -\frac{G_F} {2\sqrt{2}}
 \big\{V_{ub}^*V_{ud}[(C_1+\frac{C_2}{3})(F^{LL}_{e\rho}-F^{LL}_{a\rho}-F^{LL}_{eP}-F^{LL}_{aP})\nonumber\\
&+&C_2( M^{LL}_{e\rho}-M^{LL}_{a\rho}-M^{LL}_{eP}-M^{LL}_{aP})]-V_{tb}^*V_{td}[(C_5-\frac{C_7}{2})M^{LR}_{e\rho}\nonumber\\
&+&(\frac{7 C_3}{3}+\frac{5 C_4}{3}-2(C_5+\frac{C_6}{3})-\frac{1}{2}(C_7+\frac{C_8}{3}-\frac{2}{3}(C_9-C_{10})))F^{LL}_{e\rho}\nonumber\\
&+&(\frac{C_5}{3}+C_6-\frac{1}{2}(\frac{C_7}{3}+C_8))F^{SP}_{e\rho}+(C_3+2C_4-\frac{C_9}{2}+\frac{C_{10}}{2})M^{LL}_{e\rho}\nonumber\\
&-&(-\frac{C_3}{3}-C_4-\frac{3}{2}(C_7+\frac{C_8}{3})+\frac{5 C_9}{3}+C_{10})(F^{LL}_{a\rho}+F^{LL}_{aP})\nonumber\\
&-&(-\frac{C_5}{3}-C_6+\frac{1}{2}(\frac{C_7}{3}+C_8))(F^{SP}_{a\rho}+F^{SP}_{aP})+(2C_6+\frac{C_8}{2})M^{SP}_{e\rho}\nonumber\\
&-&(-C_3+\frac{C_9}{2}+\frac{3 C_{10}}{2})(M^{LL}_{a\rho}+M^{LL}_{eP}+M^{LL}_{aP})\nonumber\\
&-&(-C_5+\frac{C_7}{2})(M^{LR}_{a\rho}+M^{LR}_{eP}+M^{LR}_{aP})-\frac{3 C_8}{2}(M^{SP}_{a\rho}+M^{SP}_{eP}+M^{SP}_{aP})\nonumber\\
&-&(-\frac{C_3}{3}-C_4+\frac{3}{2}(C_7+\frac{C_8}{3})+\frac{5 C_9}{3}+C_{10})F^{LL}_{eP}]\big\}\;,\label{amp19}
\end{eqnarray}
 \begin{eqnarray}
{\cal A}(B^0 \to \eta_s(\rho^0\to)\pi^+ \pi^-) &=& -\frac{G_F} {2}
  \big\{-V_{tb}^*V_{td}[(C_3+\frac{C_4}{3}-C_5-\frac{C_6}{3}+\frac{1}{2}(C_7+\frac{C_8}{3}-C_9-\frac{C_{10}}{3}))F^{LL}_{e\rho}\nonumber\\
&+&(C_4-\frac{C_{10}}{2})M^{LL}_{e\rho}+(C_6-\frac{C_8}{2})M^{SP}_{e\rho}]\big\} \;,\label{amp20}
\end{eqnarray}
 \begin{eqnarray}
{\cal A}(B^0 \to \eta(\rho^0\to)\pi^+ \pi^-) &=&{\cal A}(B^0 \to \rho^0 \eta_q) \cos{\phi}-{\cal A}(B^0 \to \rho^0 \eta_s)\sin{\phi} \;,\label{amp21}\\
 {\cal A}(B^0 \to \eta^{\prime}(\rho^0\to)\pi^+ \pi^-) &=& {\cal A}(B^0 \to \rho^0 \eta_q)\sin{\phi}+{\cal A}(B^0 \to \rho^0 \eta_s) \cos{\phi} \;,\label{amp22}
\end{eqnarray}
 \begin{eqnarray}
{\cal A}(B_s^0 \to \eta_q(\rho^0\to)\pi^+ \pi^-) &=& \frac{G_F} {2\sqrt{2}}
\big\{V_{ub}^*V_{us}[(C_1+\frac{C_2}{3})(F^{LL}_{a\rho}+F^{LL}_{aP})+C_2(M^{LL}_{a\rho}+M^{LL}_{aP})]\nonumber\\
&-&V_{tb}^*V_{ts}[-\frac{3}{2}(C_7+\frac{C_8}{3}-C_9-\frac{C_{10}}{3})(F^{LL}_{a\rho}+F^{LL}_{aP})\nonumber\\
&+&\frac{3 C_8}{2}(M^{SP}_{a\rho}+M^{SP}_{aP})+\frac{3 C_{10}}{2}(M^{LL}_{a\rho}+M^{LL}_{aP})]\big\} \;
\end{eqnarray}
 \begin{eqnarray}
{\cal A}(B_s^0 \to \eta_s(\rho^0\to)\pi^+ \pi^-) &=& \frac{G_F} {2}
\big\{V_{ub}^*V_{us}[(C_1+\frac{C_2}{3})F^{LL}_{eP}+C_2M^{LL}_{eP}]-V_{tb}^*V_{ts}[\frac{3 C_8}{2}M^{SP}_{eP}\nonumber\\
&+&\frac{3}{2}(C_7+\frac{C_8}{3}+C_9+\frac{C_{10}}{3})F^{LL}_{eP}+\frac{3 C_{10}}{2}M^{LL}_{eP}]\big\} \;,\\
{\cal A}(B_s^0 \to \eta(\rho^0\to)\pi^+ \pi^-) &=& {\cal A}(B_s^0 \to \rho^0 \eta_q) \cos{\phi}-{\cal A}(B_s^0 \to \rho^0 \eta_s)\sin{\phi} \;,\\
{\cal A}(B_s^0 \to \eta^{\prime}(\rho^0\to)\pi^+ \pi^-) &=& {\cal A}(B_s^0 \to \rho^0 \eta_q)\sin{\phi}+{\cal A}(B_s^0 \to \rho^0 \eta_s) \cos{\phi}  \;,
\label{amp26}
\end{eqnarray}
where $G_F$ is the Fermi coupling constant.  $V_{ij}$'s are the Cabibbo-Kobayashi-Maskawa matrix elements.
The functions $ ( F^{LL}_{e\rho}, F^{LL}_{a\rho}, M^{LL}_{e\rho}, M^{LL}_{a\rho}, \cdots ) $
appeared in above equations are the individual decay amplitudes corresponding to different currents,
and their explicit expressions can be found in the Appendix of Ref.~\cite{Wang-2016}.


\end{document}